\documentclass{emulateapj}
\usepackage{upgreek}
\def\spitz{{\it Spitzer }}

\begin{document}

\shortauthors{Esplin \& Luhman}
\shorttitle{High-Precision IRAC Astrometry}

\title{Measuring High-Precision Astrometry with the Infrared Array Camera on the {\it Spitzer Space Telescope}\altaffilmark{1}}

\author{
T. L. Esplin\altaffilmark{2} \& 
K. L. Luhman\altaffilmark{2,3} 
}

\altaffiltext{1}{Based on observations made with the {\it Spitzer Space
Telescope}, which is operated by the Jet Propulsion Laboratory,
California Institute of Technology under a contract with NASA.}
\altaffiltext{2}{Department of Astronomy and Astrophysics, The Pennsylvania
State University, University Park, PA 16802; taran.esplin@psu.edu.}
\altaffiltext{3}{Center for Exoplanets and Habitable Worlds,
The Pennsylvania State University, University Park, PA 16802.}

\begin{abstract}
The Infrared Array Camera (IRAC) on the {\it Spitzer Space Telescope} 
currently offers the greatest potential for high-precision astrometry 
of faint mid-IR sources across arcminute-scale fields,
which would be especially valuable for measuring parallaxes of cold
brown dwarfs in the solar neighborhood
and proper motions of obscured members of nearby star-forming regions. 
To more fully realize IRAC's astrometric capabilities,
we have sought to minimize the  
largest sources of uncertainty in astrometry with its 3.6 and 4.5~$\upmu$m bands.
By comparing different routines that estimate stellar positions,   
we have found that Point Response Function (PRF) fitting with 
the {\it Spitzer} Science Center's Astronomical Point Source Extractor
produces both the smallest systematic errors from varying intra-pixel 
sensitivity and the greatest precision in measurements of positions.
In addition,
self-calibration has been used to derive new 7$^{\rm th}$ and 8$^{\rm th}$ order distortion corrections 
for the 3.6 and 4.5~$\upmu$m arrays of IRAC, respectively.
These corrections are suitable for data throughout the mission of  {\it Spitzer}
when a time-dependent scale factor is applied to the corrections. 
To illustrate the astrometric accuracy that can be achieved by combining 
PRF fitting with our new distortion corrections,
we have applied them to archival data for a nearby star-forming region,
arriving at total astrometric errors of $\sim$20 and 70 mas at signal to noise ratios of 100 and 10, respectively.
\end{abstract}


\section{Introduction}
Accurate astrometry is essential for measuring parallaxes 
for stars in the solar neighborhood ($<$30~pc) and proper motions for
members of nearby open clusters and star-forming regions (100--300~pc).
The coldest brown dwarfs near the Sun and the more obscured members of
young clusters can be difficult to detect with the optical and near-infrared
(IR) instruments that are normally used for such measurements, but they
are much brighter at mid-IR wavelengths.
The best available mid-IR instrument for astrometry of this kind
is the Infrared Array Camera \citep[IRAC;][]{faz04}
on the {\it Spitzer Space Telescope} \citep{wer04}.
IRAC has already contributed to the measurement of parallaxes for most of
the coldest known brown dwarfs
\citep{mar13,bei13,bei14,dup13,kir13,luh14b,lues14}. Meanwhile,
IRAC observations at multiple epochs spanning a decade are now available for 
several nearby star-forming regions,
making it possible to search for new members via their proper motions
(e.g., \spitz program 90071; A. Kraus).

In the two most sensitive bands of IRAC for stars and brown dwarfs
(3.6 and 4.5~$\upmu$m), 
the largest sources of uncertainty in astrometry 
are
(1) systematic offsets in the measured pixel coordinates of stars due to
varying intra-pixel sensitivity combined with an under-sampled point spread
function (PSF),
(2) $\sim$100 mas systematic errors in the \spitz pipeline's 3$^{\rm rd}$ order
distortion corrections, and
(3) random errors in the measured pixel coordinates of stars,
which can depend significantly on the algorithm adopted for measuring those
positions, especially at lower signal to noise (S/N).
To date, the first  source
has been addressed primarily  in the context of high-precision photometry of single objects
rather than astrometry of multiple sources across the IRAC arrays (see \citealt{irachpp}).
Meanwhile, 
the pipeline distortion corrections have been 
greatly improved upon with newly developed 5$^{\rm th}$ order corrections
\citep{dup13, low14},
although these new corrections are not yet publicly available. 
Finally, comparisons of algorithms that estimate stellar positions 
in IRAC images have been performed only with synthetic data \citep{mig08}
or at high S/N ratios \citep{lew13}.

In this paper, we seek to more fully realize IRAC's capabilities for 
high-precision astrometry.
We begin by describing the sets of archival IRAC data that we have 
selected for our analysis (Section 2).
Using these data,
we characterize the systematic offsets due to varying intra-pixel sensitivity
and compare the astrometric errors 
produced by several algorithms for measuring stellar positions (Section 3).
We then measure accurate, high-order distortion corrections 
for the 3.6 and 4.5~$\upmu$m bands and 
quantify the astrometric errors produced 
by combining the new corrections and our preferred 
algorithm (Section~4).

\section{Data}

The IRAC camera is described in detail by \cite{faz04}.
In summary, it contains four 256 $\times$ 256 arrays
that have plate scales of $1\farcs2$ pixel$^{-1}$,
resulting in fields of view of $5\farcm2 \times 5\farcm2$.
The arrays collect images at 3.6, 4.5, 5.8, and 8.0 $\upmu$m, which
are denoted as [3.6], [4.5], [5.8], and [8.0].
The images produced by the camera exhibit a FWHM of  $1\farcs6-1\farcs9$.
{\it Spitzer} was cooled with liquid helium from its launch in August 2003 until the depletion of the 
helium in May 2009, which is known as the cryogenic mission. It has continued
to operate in a post-cryogenic phase with the [3.6] and [4.5] bands of IRAC. 

The ideal dataset for simultaneously measuring both the distortion and 
the potential biases in algorithms for estimating stellar positions would
have the following characteristic:
it would contain a sufficiently large number of images of
moderately dense fields of stars such that each pixel in the array experiences
the detection of a star on many occasions.
Given the data available in the \spitz archive, 
we have selected post-cryogenic observations of dense fields within the Galactic Plane from
program 70157 (L. Allen).
These images contain 0.3 million stars with 3.8 million detections at S/N $>$ 15,
resulting in an average of 29 detections per pixel in each band.
To test the validity of applying the distortion corrections derived from those
data to images obtained in other years during the \spitz mission,  
we also have made use of data from 
programs 104 (T. Soifer), 146 (E. Churchwell), 20201 (E. Churchwell),
40184 (J. Hora), 60101 (R. Arendt), 70072 (B. Whitney), 80074 (B. Whitney), 
and 90071 (A. Kraus),
all of which are comparable in density of detections to the dataset
of the Galactic Plane. In addition, to compare the accuracies of 
different algorithms for measuring stellar positions,
we need observations designed to maintain point sources 
at the same positions on the array among many images.  
For this purpose, we have selected the time-series photometric monitoring
data from program 80179 (S. Metchev).

\section{Varying intra-pixel sensitivity and algorithms for measuring positions}

The sensitivity \citep[or ``gain",][]{ing12} of an individual IRAC pixel
varies from the center to the edge.
Because the PSFs in the [3.6] and [4.5] images are under sampled,
the varying sensitivity leads to correlated noise in IRAC photometry
\citep[``pixel phase effect",][references therein]{irachpp}
and systematic offsets in the measured positions \citep{mig08}.
In addition, the astrometric uncertainties can vary significantly among
different algorithms for measuring positions. 
To correct for the effect of intra-pixel sensitivity and to identify the
optimum choice of algorithm, we have compared five commonly used routines:
the IDL functions {\tt cntrd}, {\tt gcntrd}, 
and {\tt box\_centroider},
which use first derivatives, gaussian fitting, and flux-weighted means (i.e., the first moment), 
respectively;
the IRAF routine {\tt ofilter}, which employs optimum filtering and a triangular weighting 
function;
and the point response function (PRF) fitting routine implemented by the 
Astronomical Point source EXtractor \citep[APEX;][]{mak05}, which is the
extractor for the Mosaicking and Point Source Extraction package 
developed by the  \spitz Science Center (SSC).
Although location-dependent PRFs are available from the SSC,
we use only the central PRF for extraction since the use of multiple
PRFs produces discontinuities in the distortion.
We performed PRF-fitting throughout
this study using a 5~$\times$~5 pixel box around each source,
which is three times larger than the FWHM.

\begin{figure}[h]
	\centering
	\includegraphics[trim = 0mm 25mm 0mm 25mm, clip=true, scale=0.75]{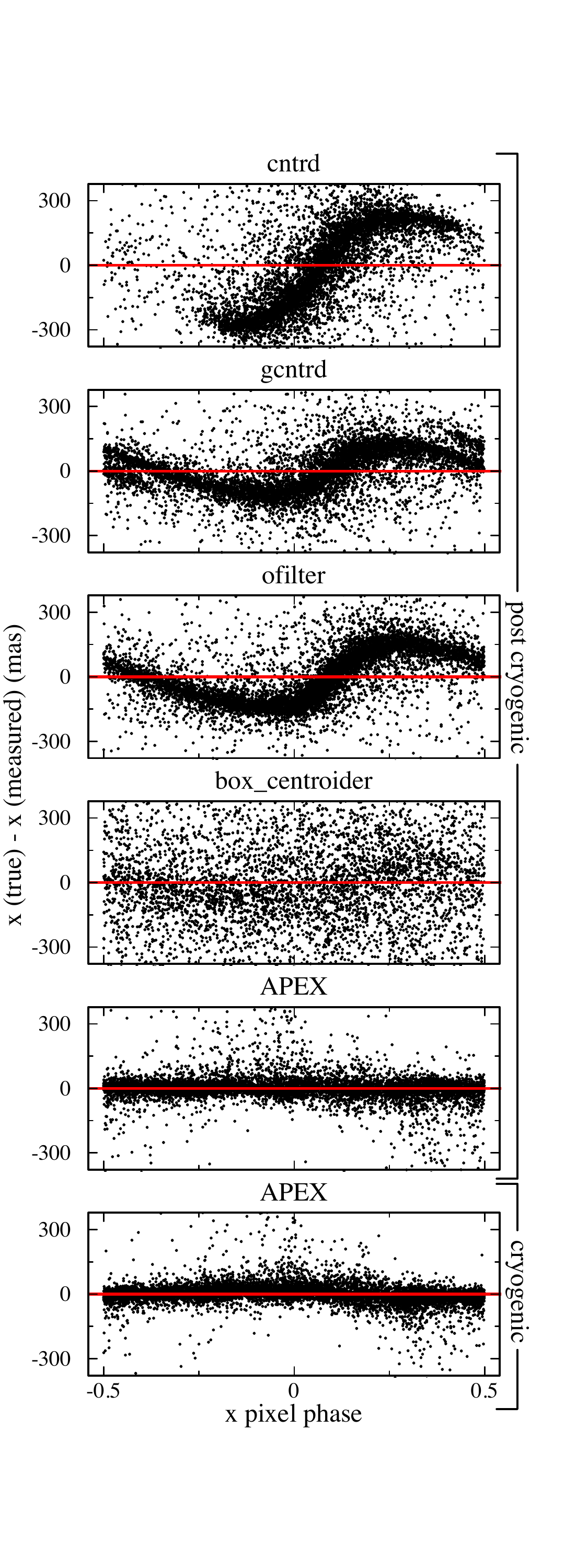}
\caption{
The differences between true and measured x positions of stars in 
[3.6] images 
 as a function distances of the measured position from the center of the pixel in the x direction 
for five routines: first derivative optimization ({\tt cntrd}),
Gaussian fitting ({\tt gcntrd}), triangular weighting ({\tt ofilter}), 
first moment ({\tt box\_centroider}), and PRF fitting (APEX).
The first four routines produce large systematic errors ($>$ 100 mas)
while APEX exhibits much smaller offsets in both the cryogenic and post-cryogenic data.
Similar results are produced for y in [3.6] and x an y in [4.5].  
Only detections with S/N $>$ 20 are plotted.
}
\label{fig:phase}
\end{figure}

\begin{figure}[h]
	\centering
	\includegraphics[trim = 0mm 0mm 0mm 0mm, clip=true, scale=0.9]{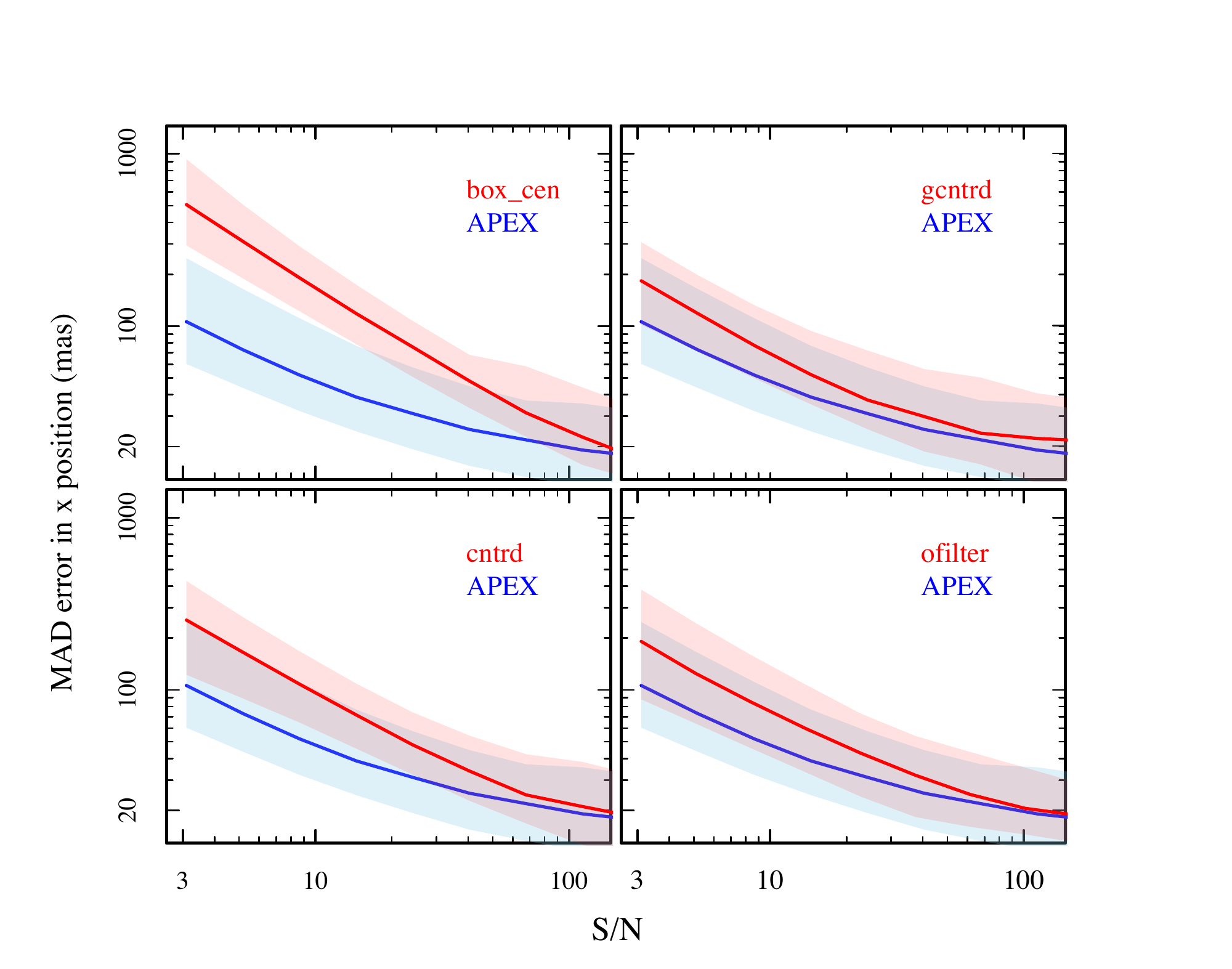}
\caption{
Comparison of [3.6] x positional errors between APEX (blue) and four other  
methods for measuring positions (red) as a function of S/N. 
The solid lines are the median errors and
the shaded regions encompass 80\% of the errors.
APEX produces the most precise positions at all values of S/N.
}
\label{fig:prec}
\end{figure}

Due to systematic offsets in the calculated positions 
between some of the algorithms (see Section 4),
we began by measuring separate preliminary distortion corrections 
with post-cryogenic data from program 70157 
for the five algorithms we considered. 
We then computed the deviations between the ``true" positions  (see Section \ref{sec:meth}) 
and the measured positions of stars. 
In Figure \ref{fig:phase},
we show the x deviations in [3.6] as a function of x pixel phase,
which is defined as the distance of the measured position from the center of the pixel 
in the x direction.
The residuals are qualitatively similar for the y pixel phase in [3.6]
and the two components in [4.5].
We find that the positions from {\tt cntrd},  {\tt gcntrd}, {\tt box\_centroider}, and {\tt ofilter}
exhibit large systematic errors ($>$ 100 mas),
which is a reflection of the varying intra-pixel sensitivity.
Those x deviations are also correlated with the y pixel phase.
For {\tt gcntrd},
we note that the offsets near the edge of the pixel form two distinct sequences  
separated by $\sim$70 mas.
The large scatter in the residuals of {\tt box\_centroider} 
is a reflection of that algorithm's low precision (see Figure~\ref{fig:prec}).

For the same post-cryogenic data analyzed with the other four
algorithms,
PRF fitting with APEX produces much smaller systematic offsets in the 
measured positions, as shown in Figure \ref{fig:phase}.
We also tested PRF fitting on the various sets of cryogenic data from Section 2,
arriving at the same result, as illustrated with the 
data from program 40184 in Figure \ref{fig:phase}.   
The [3.6] cryogenic data do show a small  
systematic offset ($\sim$10 mas),
which can be fit with a third-order function in each pixel direction (see Section \ref{sec:con}).
An offset of this kind was not found in [4.5].
It is unclear why a bias should be present only in the [3.6] cryogenic data.

To quantify the precision of the five methods for measuring positions,
we make use of the data from program 80179
because they consist of many images taken
at the same position on the sky without dithering.
For the four algorithms with large systematic errors due to the varying intra-pixel sensitivity (Figure \ref{fig:phase}), 
we applied a two-dimensional correction to the measured positions. 
We computed the median absolute deviation (MAD)
of the corrected positions for each star.
We plot the resulting MADs for the x direction in [3.6] as a function
of S/N in Figure \ref{fig:prec}.  Although all of the methods exhibit
similar precisions at high S/N ($\sim$20 mas at S/N = 100), PRF fitting
 produces the smallest errors at lower S/N ($\sim$100 mas at S/N = 3).

Because PRF fitting with APEX produces the smallest systematic offsets
due to intra-pixel sensitivity variations and the highest precision,
we used it for deriving our distortion correction,
and recommend its use when precision IRAC astrometry is required. However,
{\tt ofilter} is a reasonable alternative since its systematic errors 
can be corrected and its precision is only slightly worse than that of APEX. 

\section{Distortion Correction}

\subsection{Formalism for the Distortion Correction}
\label{sec:con}
We define our distortion correction in a way that is compatible with   
the Simple Imaging Polynomial (SIP) convention
(\citealt{shu05}),
which was developed for \spitz early in its mission.
Following the syntax of \cite{cal02},
we define $u'$ and $v'$
to be the positions measured by APEX after subtracting 128.
 $u'$ and $v'$
are converted to world coordinates (right ascension and declination)
in the following manner:
\begin{enumerate}
\item If the positions are measured from [3.6] cryogenic images, $u'$ and $v'$ are 
corrected for varying intra-pixel sensitivity as follows:
\begin{eqnarray}
u = u' + f'(u' - {\rm floor}(u')) \\
v = v' + g'(v' - {\rm floor}(v'))
\end{eqnarray}
where $f'$ and $g'$ are defined as
\begin{eqnarray}
f'(u') = \sum_{p} A_{p} u'^p, \  0 \leq p \leq {\rm order} \\
g'(v') = \sum_{p} B_{p} v'^p, \ 0 \leq p \leq {\rm order}.
\end{eqnarray}
The values of $A_{p}$ and $B_{p}$ are provided in 
an R package described at the end of Section 4.3.
As discussed in Section 3,
positions from [3.6] post-cryogenic images and [4.5] images do not require such
corrections (i.e., $u' = u$ and $v' = v$).

\item $u$ and $v$ are converted to intermediate 
coordinates $x$ and $y$ by
\begin{equation}
\left( \begin{array}{c}
x \\
y  \end{array} \right) = 
\left( \begin{array}{rr}
\cos\theta & - \sin\theta \\
\sin\theta & \cos\theta \end{array} \right)
\left( \begin{array}{c}
s(t) * f(u,v) \\
s(t) * g(u,v) \end{array} \right)
\end{equation}
where $\theta$ is the position angle of the y axis, 
$s$ is a time-dependent scale factor applied to the distortion correction (see Section \ref{sec:time}),
and $f$ and $g$ contain the distortion correction defined in the following 
manner:
\begin{eqnarray}
f(u,v) = \sum_{p,q} A_{pq} u^p v^q, \  1 \leq p+q \leq {\rm order}_x \\
g(u,v) = \sum_{p,q} B_{pq} u^p v^q, \ 1 \leq p+q \leq {\rm order}_y.
\end{eqnarray}
As with  $A_{p}$ and $B_{p}$, 
the values of $A_{pq}$ and $B_{pq}$ are found in the R package
in Section 4.3.

\item Finally, world coordinates are computed by following the gnomonic 
projection as described by \cite{cal02},
which takes $x$, $y$, and a frame's central world coordinates as input.

\end{enumerate}

We also use the inverse of these steps 
(i.e., the conversion from world coordinates into any of the intermediate steps)
in our measurement  of $A_{p}$, $B_{p}$, $A_{pq}$, and $B_{pq}$. 
The second and third steps have analytic inversions,
while the distortion corrections themselves do not, which 
introduces an insignificant error ($<$0.01 pixel) to the conversion.

\subsection{Measurement of the Distortion Correction}
\label{sec:meth}

To measure distortion corrections with the data from program 70157 (Section 2),
we could either adopt an independent catalog of high-precision astrometry 
or apply self calibration 
(i.e., simultaneously measure an astrometric catalog, distortion corrections of [3.6] and [4.5], and relative offsets
and orientations among images).
We choose to employ the latter because any independent catalog would have different resolution
and sensitivity than the IRAC images.
Because self-calibration involves millions of parameters that depend on each other nonlinearly, 
we derived initial distortion corrections and measured 
relative offsets and orientations using astrometry from
the Two Micron Point Source Catalog \citep[2MASS,][]{skr06}
and then we refined the distortion corrections iteratively
in the following manner.
(1) We constructed an astrometric catalog 
by calculating the mean position for each star 
from individual detections with S/N $>$ 15
using our current distortion correction and relative offsets and orientations.
(We refer to the mean position as a star's ``true" position,
and the detections as the star's ``measured" positions.)
To quantify the goodness of fit for the current corrections and offsets,
we  calculated the RMS of all measured positions relative to the true positions.  
(2) We alternated solving for new distortion corrections
and the relative offsets using  least-squares algorithms defined in R \citep[an open source statistical software package;][]{R}.
(3) We continued iterating these two steps until the RMS declined to a roughly constant value. 
In the second step,  
after applying the appropriate partial inverse transformations  to the true positions,
we used the linear modeling function {\tt lm} in R to evaluate the distortion-corrections coefficients.
Meanwhile, we measured the relative orientations and offsets between frames using the nonlinear R
algorithm {\tt optim}. 
After completing an initial series of iterations and converging on a solution, 
we identified and removed detections on bad pixels 
and on the edges of the arrays where APEX produces erroneous positions 
and then repeated the iterative process. 

\begin{figure}[h]
	\centering
	\includegraphics[trim = 0mm 0mm 0mm 0mm, clip=true, scale=0.7]{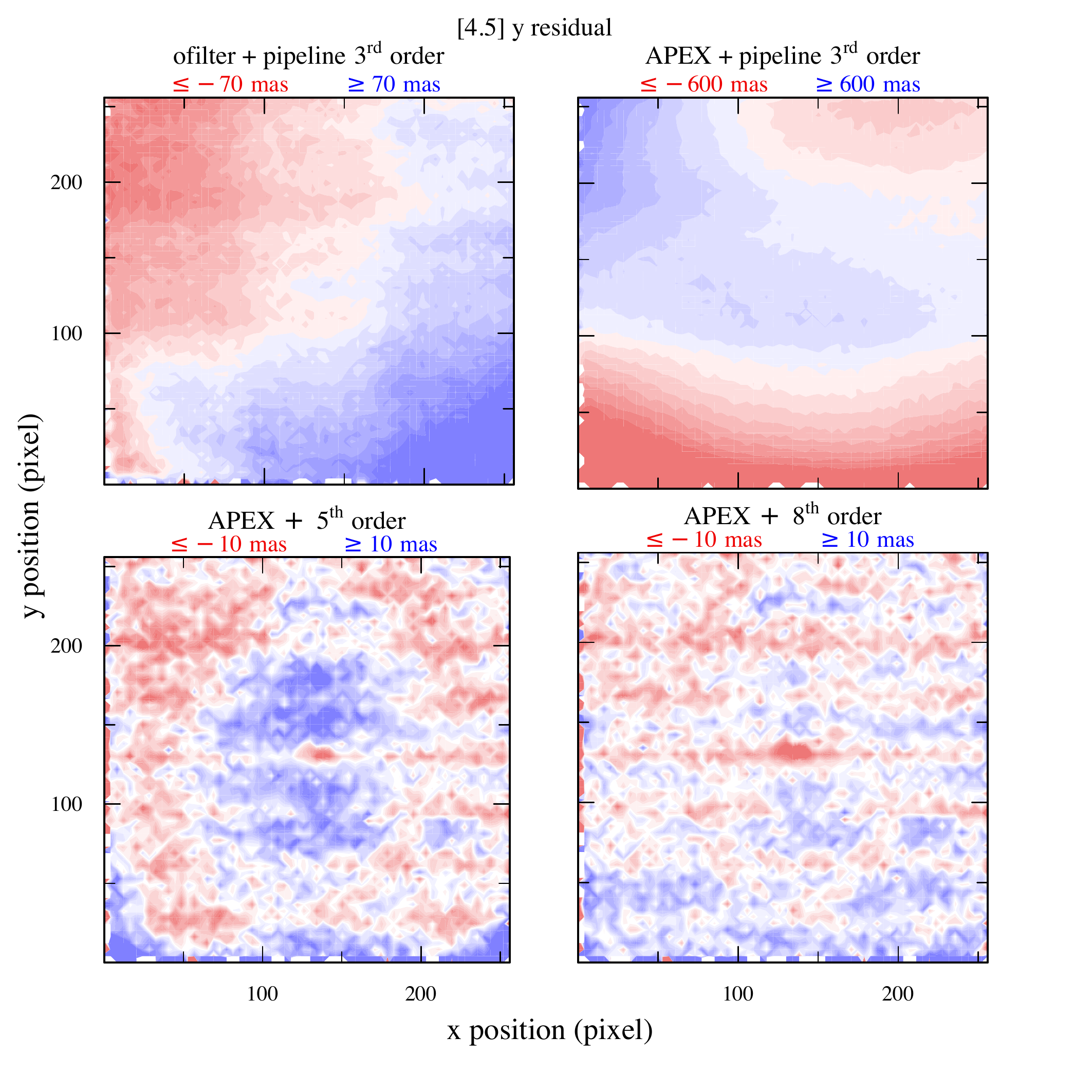}
\caption{
Deviations from true y positions of sources in [4.5] images for different
combinations of algorithms for measuring pixel coordinates and distortion
corrections.  The correction in the top panels is from the \spitz data
pipeline while those in the bottom panels have been derived in this work.
In the top panels, we find that the {\tt ofilter} routine produces much
smaller residuals than APEX for the pipeline distortion correction.
The bottom panels demonstrate that higher order distortion corrections can
produce more accurate astrometry than the 3$^{\rm rd}$ order pipeline
correction.
}
\label{fig:comp}
\end{figure}

To determine the optimum orders for equations (6) and (7),
we considered a wide range of orders (3 to $>$20) when analyzing the data from program 70157.
We then applied the solution derived for each order 
to the subset of observations from program 90071 
that encompassed the Taurus star-forming region.
For comparison,
we also applied the \spitz pipeline's 3$^{\rm rd}$ order corrections
to those data as well.
We measured the relative orientations and offsets between images 
using the same self-calibration method as before,
except with the distortion corrections fixed.
To quantify the errors for each set of distortion corrections, 
we calculated the median deviation of measured positions from 
true positions in Taurus data for 4$\times$4 pixel bins across the [3.6] and [4.5] arrays.
In Figure \ref{fig:comp},
we show the median deviations in the y direction 
of the [4.5] array produced by (1) the pipeline correction and {\tt ofilter} positions,
(2) the pipeline correction and APEX positions,
(3) our 5$^{\rm th}$ order correction and APEX positions,
and 
(4) our 8$^{\rm th}$ order correction and APEX positions. 
A time-dependent scale factor described in Section \ref{sec:time}
was applied to our distortion corrections. 
The pipeline correction produces residuals ranging from $-$80 to 120 mas
and $-$1000 to 450 mas
when applied to {\tt ofilter} and APEX positions, respectively.
The difference between the residuals produced by the two algorithms 
is a reflection of offsets in the measured pixel positions between APEX and
{\tt ofilter}, which are likely due to variations of the PSF as a function
of array location. 
This comparison demonstrates 
that the systematic astrometric errors can depend significantly on 
the adopted algorithm used to measure positions for a given distortion correction. 
{\it Indeed, our distortion corrections are only valid when used in conjunction with APEX PRF fitting.}

\begin{figure}[h]
	\centering
	\includegraphics[trim = 0mm 0mm 0mm 0mm, clip=true, scale=0.7]{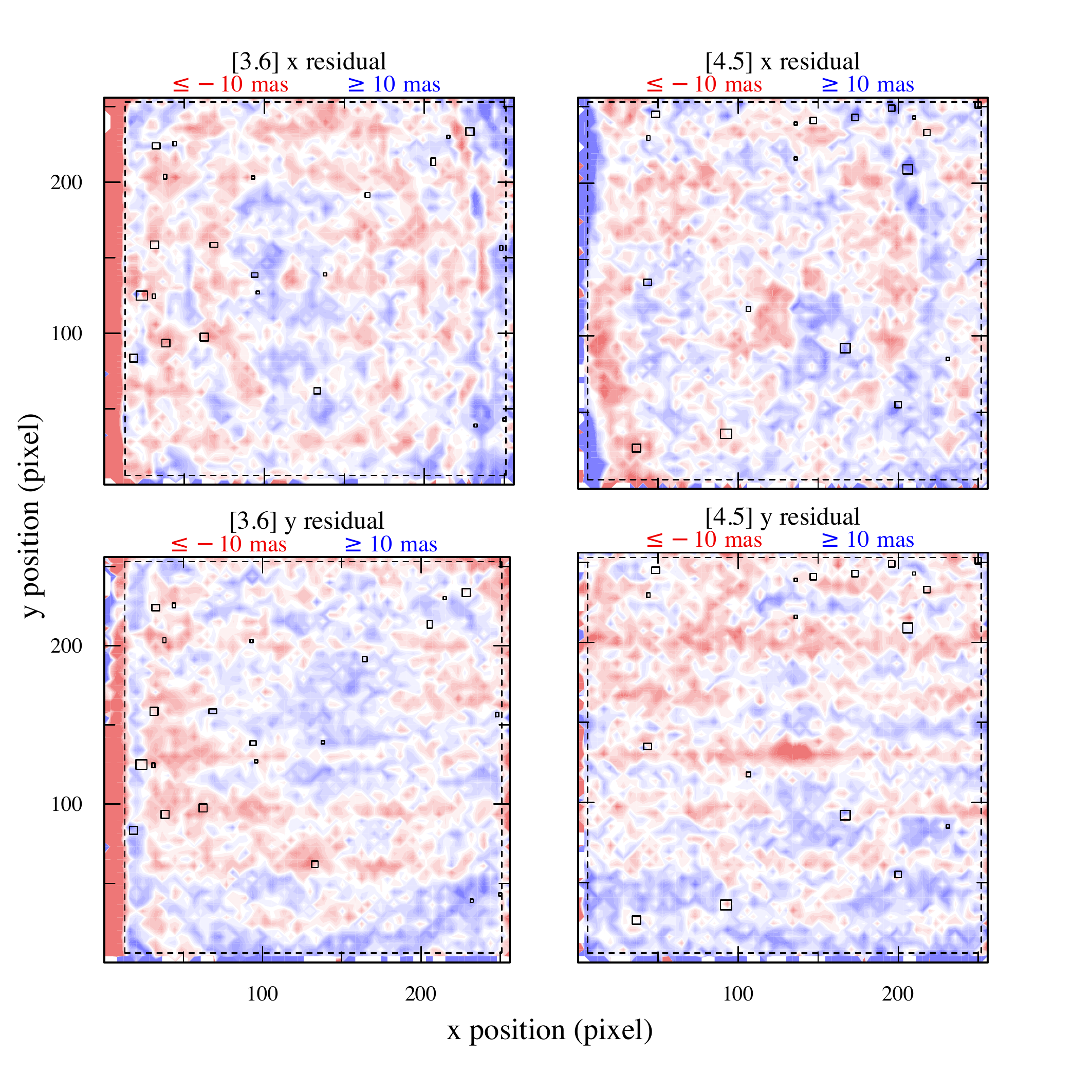}
\caption{
Deviations from true x and y positions of sources in  [3.6] and [4.5] images 
of Taurus produced by our final 7$^{\rm th}$ and 8$^{\rm th}$ order corrections,
which were derived from data in the Galactic plane. 
We mark bad pixels (within solid rectangles) and pixels near the edges 
of the arrays (outside of dotted lines)
where APEX produces erroneous positions,
which were omitted from the derivation of our distortion corrections.
}
\label{fig:resids}
\end{figure}

The Bayesian information criterion \citep{sch78}
indicated an optimum order of $\sim$20 for our distortion corrections
when applied to the dataset used to measure the distortion.
However, when we examined residual maps like those in the top panel
in Figure \ref{fig:comp}
for different orders on an independent dataset, we found that the improvements
beyond orders of $\sim$7 for [3.6] and $\sim$8 for [4.5] were much smaller
than the minimum errors in the measurements of pixel coordinates
($\sim$20~mas, Fig.~\ref{fig:prec}).
Therefore, we adopted 7$^{\rm th}$ order corrections for 
[3.6] and 8$^{\rm th}$ order corrections for [4.5].
To illustrate the improvement of  8$^{\rm th}$ order relative to lower orders,
we show in Figure \ref{fig:comp} the residual maps using 5$^{\rm th}$ and  8$^{\rm th}$ order corrections.
In Figure \ref{fig:resids}, we plot the median x and y deviations 
for both bands in the Taurus data produced by our adopted corrections.
The large residuals near the edges in Figure \ref{fig:resids} are
due to the fact that they were omitted from the fitting process because of the
erroneous positions from APEX at those locations, as noted earlier.

\subsection{Time Dependence of the Plate Scale}
\label{sec:time}

\begin{figure}[h]
	\centering
	\includegraphics[trim = 25mm 0mm 0mm 0mm, clip=true, scale=0.75]{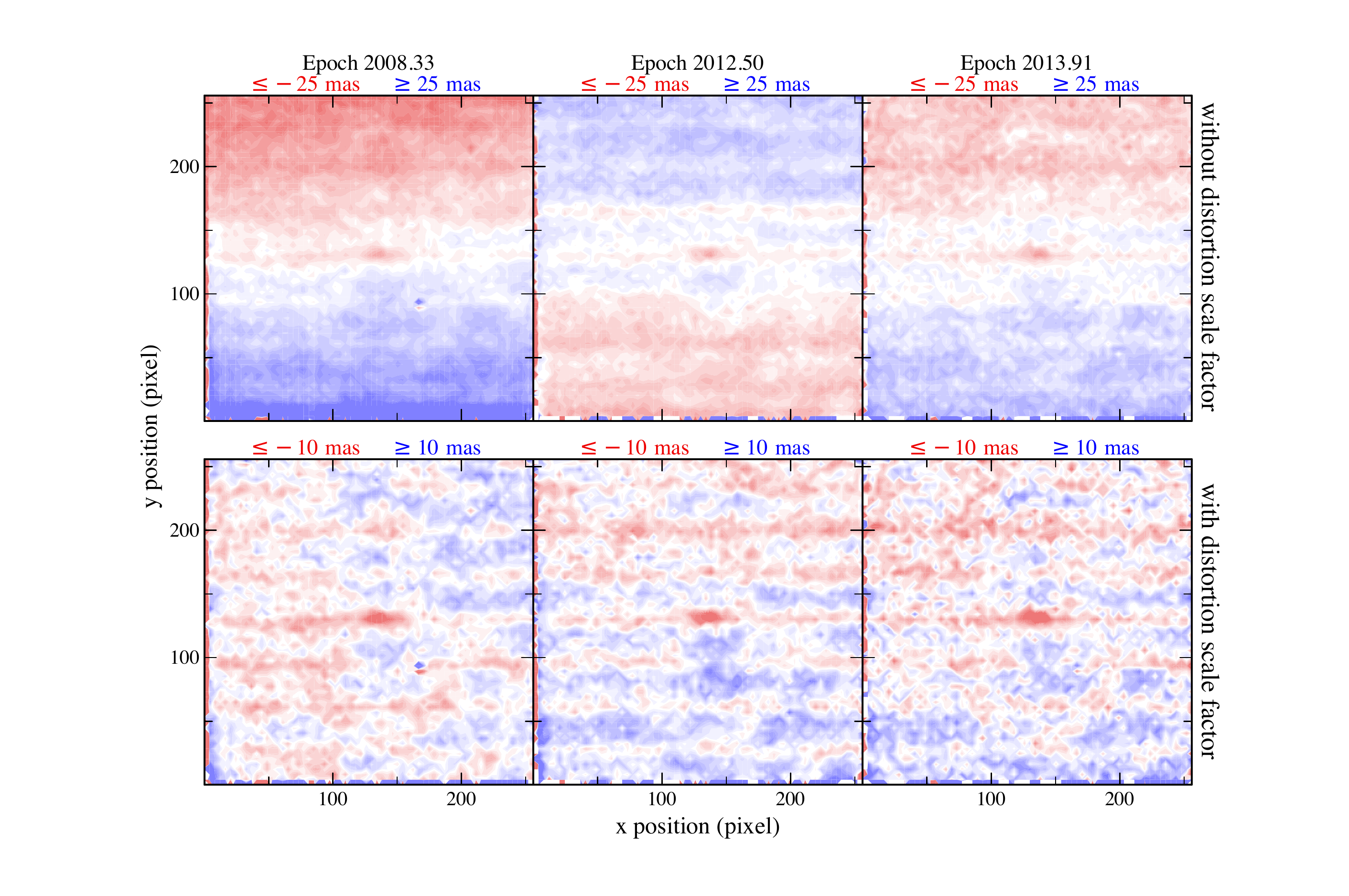}
\caption{
Deviations from true y positions of sources in [4.5] images for 
one cryogenic epoch (2008.33) and two post cryogenic epochs (2012.50 and 2013.91).
The residual maps in the top panel are produced with our distortion correction derived from 
data in the Galactic Plane (2011.58).
The systematic offsets in each of those maps can be fit by a scale factor applied 
to the distortion correction.
After applying such scale factors, we arrive at the residual maps in the 
bottom panel.
The scale factors that we derive from these maps and datasets at additional epochs
are shown in Figure \ref{fig:scale}. 
}
\label{fig:scalecorec}
\end{figure}

\begin{figure}[h]
	\centering
	\includegraphics[trim = 10mm 0mm 0mm 0mm, clip=true, scale=1]{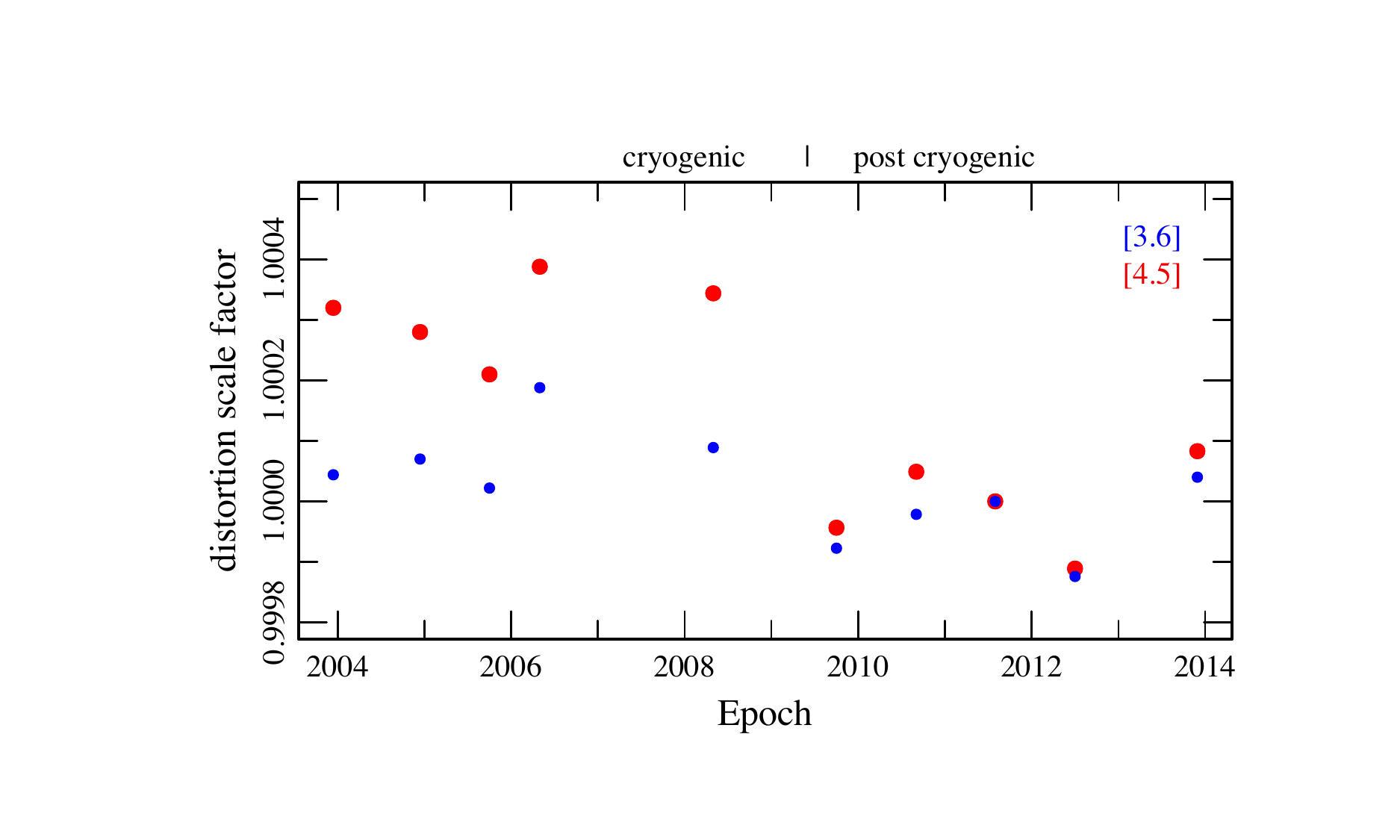}
\caption{
Scale factors applied to our distortion corrections for [3.6] and [4.5]
for datasets from various epochs during the {\it Spitzer} mission.
The application of these factors removes the systematic offsets in residual maps like those shown in Figure 5.
}
\label{fig:scale}
\end{figure}

\begin{figure}[h]
	\centering
	\includegraphics[trim = 0mm 0mm 0mm 0mm, clip=true, scale=0.9]{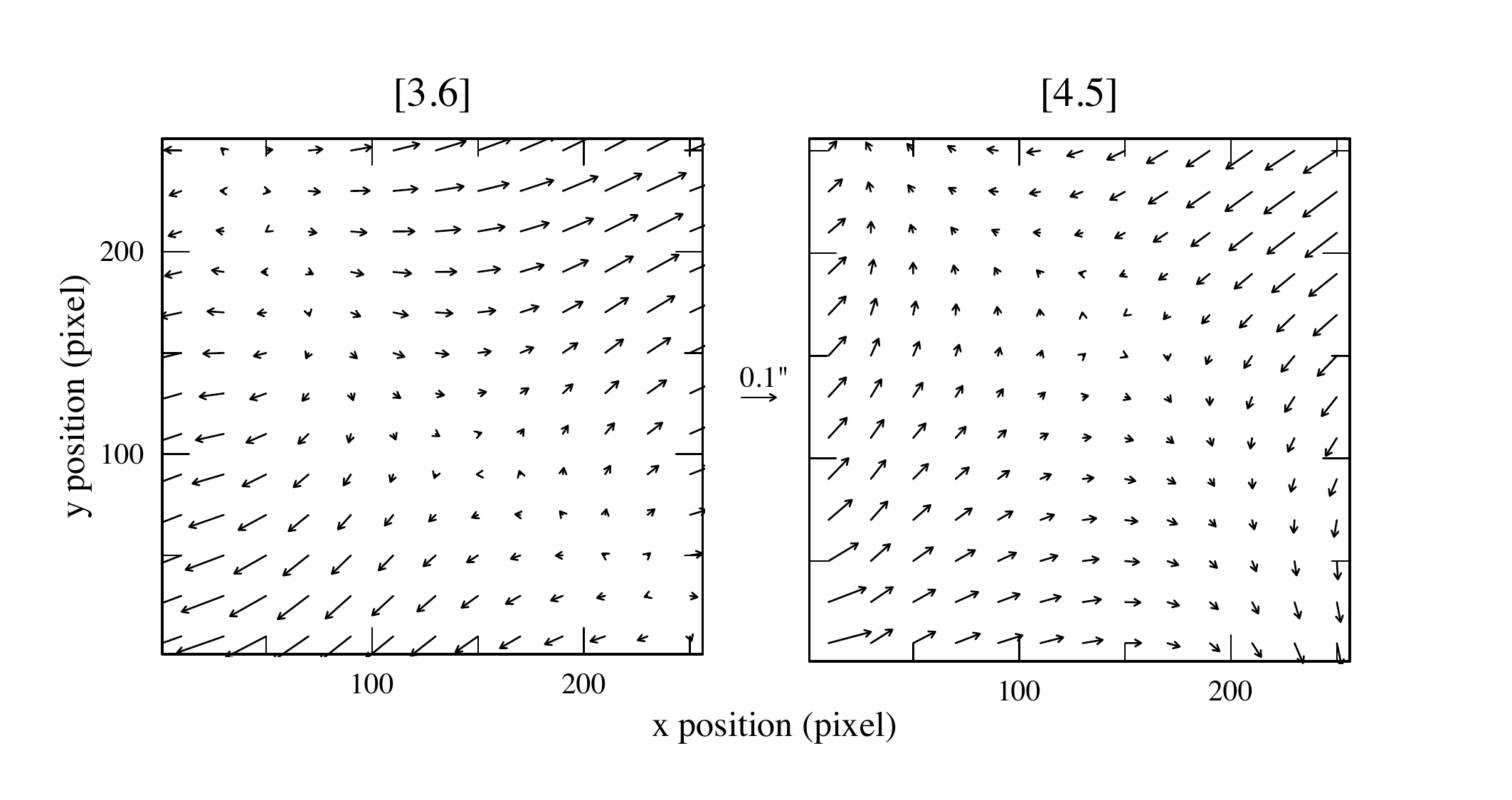}
\caption{
Deviations of the distortions implied by our 7$^{\rm th}$ and 8$^{\rm th}$
corrections from a uniform plate scale
for [3.6] and [4.5].  
The deviations have been magnified by a factor of 200.  
}
\label{fig:final}
\end{figure}

We investigated whether our adopted distortion corrections
are valid throughout the \spitz mission
by applying them to the datasets selected in Section 2.
For each dataset,
we found a systematic low-order offset in the residual maps.
As an illustration, in the top panel of Figure \ref{fig:scalecorec}
we show the residual maps for the y component of [4.5]
for programs 20201, 80074, and 90071,
which have epochs of 2008.33, 2012.50, and 2013.91, respectively.
At each epoch, there are two or three sections of the array with 
distinctly different median residuals, which range from about $-$25 to 25 mas.
Similar trends in the x direction are seen in the residual maps of the x component of [4.5].
The [3.6] residual maps also show offsets of this kind, except with a smaller
range. This pattern is likely a reflection of the tiling of images in a
given dataset. We find that these offsets can be can be eliminated by 
multiplying the distortion correction by a scale factor,
as demonstrated in the bottom panel of Figure \ref{fig:scalecorec}.
We plot these scale factors for all of the datasets that we have analyzed 
as a function of epoch in Figure \ref{fig:scale}.
The value of the factor appears to vary on a time scale of several months 
with the largest change occurring at the transition between the cryogenic
and post-cryogenic phases of the mission. 
The sizes of the scale factors are equivalent to changes of  a few millimeters 
in the focal length of the telescope. The scale factor for 
an epoch not represented in our selected datasets  
can be interpolated from the scale factors in Figure \ref{fig:scale}
or measured independently in the way that we have done
if the dataset is large enough. In Figure \ref{fig:final},
we show maps of the magnitude and direction of the distortion implied by
our adopted corrections for [3.6] and [4.5] at epoch 2013.91.

Because the number of coefficients within a single distortion correction is
high ($\sim$70) and our corrections were measured with R,
we have created an R package ``IRACpm"\footnote{Available at 
GitHub: \url{https://github.com/esplint/IRACpm}}
that contains functions and instructions for applying the corrections.
Although the formalism of Section 4.1 is compatible with SIP,
we do not recommend simply updating the headers of \spitz images 
with our distortion correction because
[3.6] cryogenic images benefit from 
a correction for varying intra-pixel sensitivity
prior to the application of the distortion and 
our corrections are only applicable  
to positions measured by APEX (see Figure \ref{fig:comp}).

\subsection{Application to an Example Dataset}

\begin{figure}[h]
	\centering
	\includegraphics[trim = 0mm 0mm 0mm 0mm, clip=true, scale=0.7]{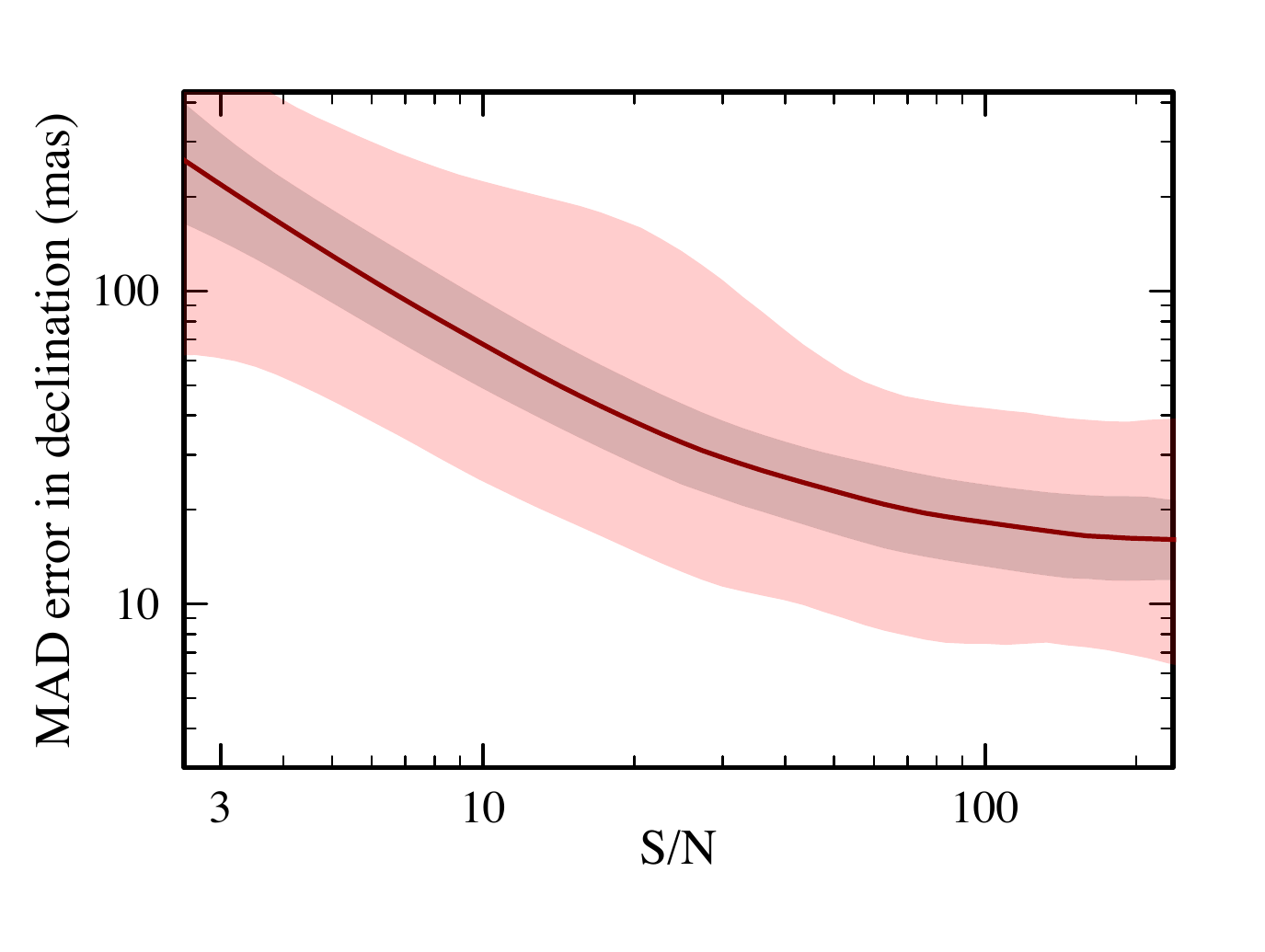}
\caption{
Errors in declination for sources in IRAC images of the Chamaeleon I 
star-forming region as a function of S/N after applying our adopted distortion corrections
(grey points). 
The median errors are plotted as the sold line
and the darker and lighter shaded regions encompasses 50\% and 90\% of the errors,
respectively.
At S/N $>$ 100, the errors are less than 20 mas for half of the sources, 
which is the expected value if the error was due almost entirely to the measured positions 
(see Fig.\ \ref{fig:prec})
rather than errors in the distortion corrections or the relative positions and orientations among the 
dithered images.
}
\label{fig:cham}
\end{figure}

To illustrate the astrometric accuracy that can be achieved by using
our new distortion corrections,
we applied them to archival multi-epoch images of  the 
Chamaeleon I star-forming region (T. Esplin, in preparation).
We present the MADs as a function of S/N for 
the final epoch of observations (program 90071) in Figure \ref{fig:cham}.
The median positional errors are $\sim$70~mas at S/N = 10 
(90\% of the errors are between 25 and 230~mas)
and $\sim$20~mas at S/N = 100 (90\% between 7 and 42~mas).
Most of the error at low S/N is likely introduced in the measurement of the  
pixel coordinates of a star (see Fig.~\ref{fig:prec}) with only minor
contributions from the relative offsets and orientations between images
and the distortion corrections.

\section{Conclusions}

We have sought to characterize and correct for the largest sources of
uncertainty in astrometry in the [3.6] and [4.5] bands of IRAC.
We began by comparing five commonly used routines for measuring
the pixel coordinates of stars that implement flux-weighted means,
Gaussian fitting, triangular weighting, first derivatives, and PRF fitting.
We have found that PRF fitting with APEX produces the 
smallest systematic errors from varying intra-pixel sensitivity.
After correcting for those systematic errors for each algorithm,
APEX provides the greatest precision in the pixel coordinates of
sources at all values of S/N.  For these reasons, 
we recommend the use of PRF fitting with APEX for IRAC analysis that requires 
precise astrometry (e.g., photometric monitoring for transits or variability
and proper motion surveys).

By applying self-calibration to IRAC data in the Galactic plane,
we have derived 7${\rm th}$ and 8${\rm th}$ order 
distortion corrections for the [3.6] and [4.5] arrays, respectively.
To test these distortion corrections, we have applied 
them to datasets spanning the mission of {\it Spitzer}.
The resulting residual maps contain low-order systematic offsets 
that are well-fit by a time-dependent scale factor that
is multiplied by the distortion correction.
After correcting for this effect,
these datasets exhibited residuals that have a range of roughly $\pm$10 mas,
which reflects the accuracy of our distortion corrections.
When analyzing data in the Chamaeleon I star-forming region by 
combining APEX PRF fitting with our new distortion corrections,
we have achieved total astrometric errors of $\sim$20 and 70 mas at S/N = 10 and 100, respectively.
To promote the use of IRAC for precision astrometry,
we have made our distortion corrections publicly available in
a new R package called ``IRACpm". 
We note that our corrections are only valid when used in conjunction
with positions measured with APEX PRF fitting.

\acknowledgements
We acknowledge support
from grant AST-1208239 from the National Science Foundation.
We thank Patrick Lowrance and John Stauffer for helpful
discussions regarding the distortion correction for IRAC.
2MASS is a joint project of the University of Massachusetts and the Infrared Processing and Analysis
Center at Caltech, funded by NASA and the NSF.
The Center for Exoplanets and Habitable Worlds is supported by the
Pennsylvania State University, the Eberly College of Science, and the
Pennsylvania Space Grant Consortium.

\clearpage

\end{document}